\documentclass[prb,reprint]{revtex4-1}

\usepackage{graphicx}
\usepackage{epstopdf}
\usepackage{amsmath}
\usepackage{amssymb}
\usepackage{mathrsfs}

\begin{document}

\title{Magnetic reversal modes in multisegmented nanowire arrays with long aspect ratio}
\author{E. A. Rando}
\author{S. Allende}

\affiliation{Departamento de F\'\i sica, CEDENNA, Universidad de
Santiago de Chile, USACH, Av. Ecuador 3493, Santiago, Chile}

\date{\today }

\begin{abstract}
A detailed numerical analysis of the magnetization reversal processes in multisegmented nanowire arrays was developed. The nanowires have a long aspect ratio and are formed by magnetic and non-magnetic sections alternately arranged in such a way that the array resembles magnetic layers separated by non-magnetic layers.  Attention has been focused on the influence of magnetostatic interaction in the magnetic pattern formation of these magnetic nanostructures. Results from a magnetic correlation function among layers show that three different reversal modes can be detected depending on the number and distance between the magnetic segments. As a consequence of the different reversal modes, a non-monotonic behavior of the annihilation field in function of the distance between the layers is evidenced. Thus, these results are important for the production of magnetic devices with multisegmented nanowire arrays.\end{abstract}

%\pacs{75.60.Ch, 75.60.Ej, 75.60.Jk}

\maketitle

\section{Introduction}
\indent Magnetic nanoparticles have been extensively investigated due to their potential uses in technological applications, such as magnetic data storage, spintronic devices and magnetic sensors\cite{Abramovich20}. In particular, magnetic cylindrical nanowires are good candidates due to their magnetic stability and the possibility to fabricate them in a hexagonal array with long-range order and well-controlled geometry. The hexagonal array of nanowires can be obtained by filling the alumina pores with magnetic material by electrodeposition process. Furthermore, the synthesis of arrays of multisegmented nanowires with precise control over segment dimensions is available by electrodeposition method, alternating magnetic and nonmagnetic materials during the fabrication process\cite{fert,wan,lee,Bangar,Nasirpouri,addref,garcia}.  

Multisegmented elements have been proposed for device storage, because they can store more than one bit depending on the number of elements and exhibit large magnetic stability of their states\cite{albrech,escrig2,boris}. For instance, Cisternas and Vogel studied the stability of ferromagnetic patterns inscribed on arrays of multisegmented magnetic nanocylinders\cite{vogelmulti}. The systems were hexagonal arrays of multisegmented nanowires where every pore had two, three or four magnetic segments. They found that these multisegmented nanowire arrays are more appropriate to inscribe ferromagnetic patterns and store fixed information than nanowire arrays due to the reduction of the magnetostatic energy per nanowires \cite{vogelmulti,vogelnanohilo}.  With the purpose to use these systems for magnetic devices, it is necessary to understand how their magnetization reverts in those nanostructures. Allende \textit{et al.} observed two reversal modes for a single multisegmented nanowire at some constant applied field\cite{allende}. These reversal modes are serial or parallel and depend on the distance between the segments. But when the systems are formed with several multisegmented nanowires, the reversion of these systems becomes more complex due to the magnetostatic interaction among them.

In this work we investigate how the magnetostatic interaction affects a system of multisegmented magnetic nanowires with long aspect ratio by changing the number of magnetic segments by pores, as well as the spacing between the segments of the same pore, using Monte Carlo simulation. In addition, we study the nucleation and annihilation fields in different hysteresis curves. Further, we analyze the magnetic correlations among magnetic layers for these systems, and we observe three different reversal modes for the magnetization of these systems.

\section{Theory}

\begin{figure}[htb]
\centering
\includegraphics[scale=0.3]{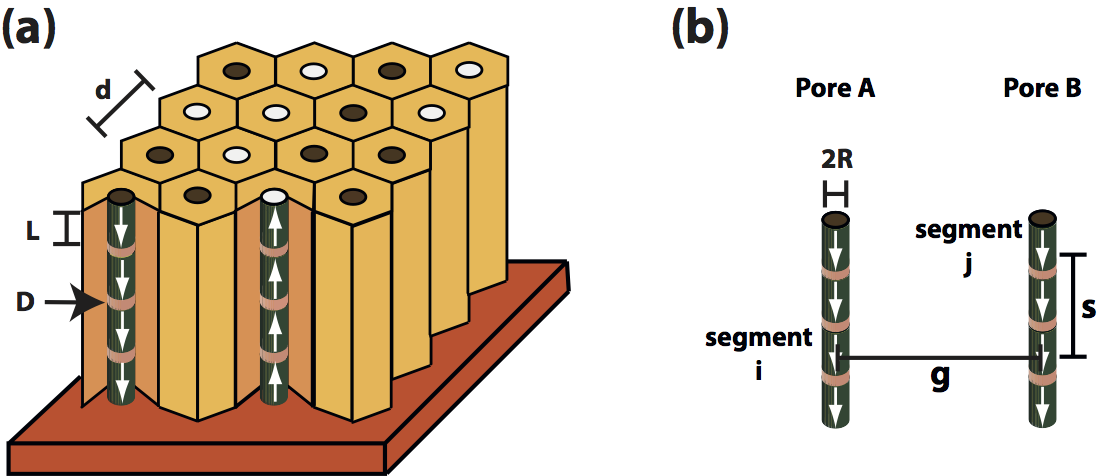}
\caption{(Color online) Schematic representation of a multisegmented nanowire array with $N_L=4$ (a). Geometrical parameters (b). }
\label{fig1}
\end{figure}

Our starting point is a system formed by an alumina matrix with dimension of $156.25$ $\mu$m$^2$, pore radius of $R=90$ nm, and pore separation of $d=500$ nm. The total number of pores in the alumina is $N_P=740$. Every pore is filled with $N_L$ magnetic Ni segments, of radius $R=90$ nm and length $L=3600$ nm, separated by a non-magnetic material with dimension $D\geq 10$ nm, see figure \ref{fig1}. The non-magnetic spacer is metallic, for instance, Cu, which is thick enough to avoid exchange and Ruderman-Kittel-Kasuya-Yosida (RKKY) interactions.

 In order to study the hysteresis loop and the magnetic reversal of these systems, we use Monte Carlo simulation considering magnetostatic interactions among the nanowires\cite{escrig}. For our simulations, we have considered the geometrical parameters ($R=90$ nm, $L=3600$ nm and $d=500$ nm) because we can do the following approximation. We consider that every segment has two magnetic stable states along the $z$ direction (up and down), because they have a strong shape anisotropy due to the large aspect ratio of them, $L/(2R)=20$\cite{sampaio,escrig}. The internal energy can be written as $E_{tot}=-\mu_0 M_0 V \sum_{i=1}^N \vec{\sigma}_i \cdot \vec{H}_{eff}$, where N is the total number of segments, $M_0=480$ emu$/$cm$^3$ is the Ni saturation magnetization, $\vec{\sigma}_i=\pm \hat{z}$ represents the stable magnetic states for the $i$ segment, and  $\vec{H}_{eff}$ is the effective field given by
 
 \begin{equation}\label{Heff}
\vec{H}_{eff}=\vec{H}+\vec{H}_{a,i}-\frac{1}{2} \sum_{\substack{
  j=1\\
j \neq i}}^N D_{ij} \vec{\sigma}_j.
\end{equation}
 
The first term of the effective field, $\vec{H}$, corresponds to the external applied field fixed along the $z$ direction of the structure. The second term, $\vec{H}_{a,i}$ , is the magnetic shape anisotropy field of an isolated segment. In our case, $|\vec{H}_{a,i}|=0.228$ kOe\cite{escrig}. The magnetic reversal mode of a single segment has been considered with the value of this parameter. The last term in equation 1 represents the magnetostatic field. The coupling constant $D_{ij}$ has been calculated by Escrig \textit{et al.}\cite{escrig1} and is given by

\scriptsize
\begin{equation}
D_{ij} =  \frac{M_0R^2}{4L}\left(\frac{2}{\sqrt{g^2+s^2}}-\frac{1}{\sqrt{g^2+(L-s)^2}}-\frac{1}{\sqrt{g^2+(L+s)^2}}\right) 
\label{Etot}
\end{equation}

\normalsize

where $s$ is the vertical and $g$ the horizontal distances between the segments $i$ and $j$, see figure 1(b).

\begin{figure}[htb]
\centering
\includegraphics[width=8.5cm,height=15.0cm]{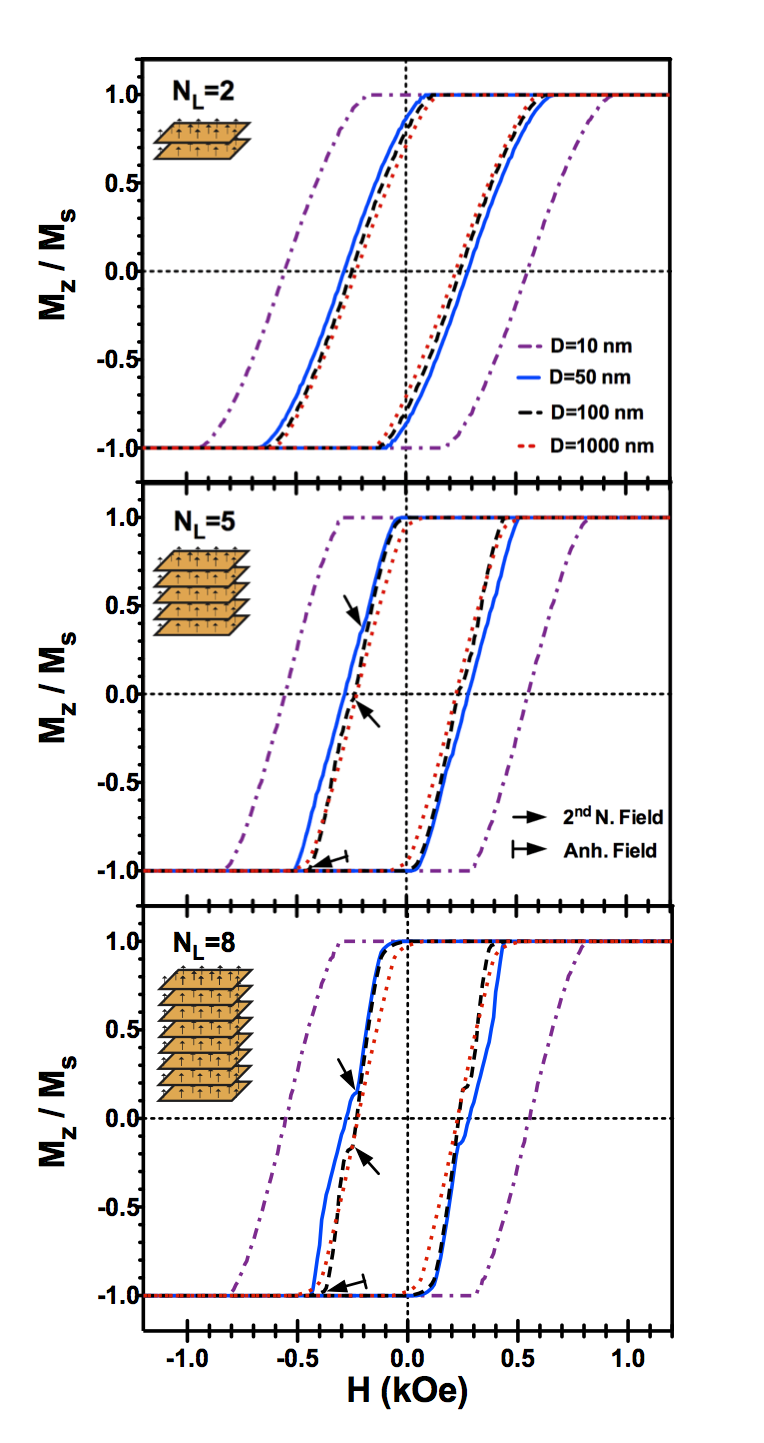}
\caption{(Color online) The figure illustrates the hysteresis curves for (a) $N_L=2$, (b) $N_L=5$, and (c) $N_L=8$ at distance between layers  of D= $10$, $50$, $100$ and $1000$ nm. The black arrow from bar, $\mapsto$, shows the annihilation field. The black arrow, $\rightarrow$, shows the second nucleation field. }
\label{fig2}
\end{figure}

Monte Carlo simulations were carried out at $T=300$ K using regular Metropolis algorithm\cite{Binder}. The new orientation of the variable $\vec{\sigma}_i$ was chosen arbitrarily with a probability $p =$ min$\left[1,\exp \left(-\Delta E/k_BT\right)\right]$, where $\Delta E$ is the change in energy due to the reorientation of  $\vec{\sigma}_i$ and $k_B$ is the Boltzmann constant. If the new orientation is accepted from $\vec{\sigma}_i$ to $-\vec{\sigma}_i$, the shape anisotropy field must change from  $\vec{H}_{a,i}$ to $-\vec{H}_{a,i}$. The shape anisotropy field for an isolated segment changes its sign after reverting its magnetization, because the new magnetic state is a stable magnetic state. The hysteresis curves with normalized magnetization start with a field higher than the saturation field, $1.5$ kOe, where the initial state for all the segment is $\vec{\sigma}_i=\hat{z}$ and $\vec{H}_{a,i}=0.228$ $\hat{z}$ kOe. The field was then linearly decreased at a rate of 300 Monte Carlo steps for $\Delta H=0.01$ kOe. The values of nucleation and annihilation fields correspond to an average over, at least, 5 independent realizations. 

\section{Results and Discussions}

The multisegmented nanowire system we have considered consists of N segments distributed in $N_L$ layers, where 740 segments are in every layer (the system with $N_L$ segments in one alumina pore is equivalent to have $N_L$ layers with one segment). The study was done for $N_L=1,...,13$ layers with distance between them of $D=10$, $20$, $50$, $100$, $250$, $500$, and $1000$ nm, respectively.

\begin{figure}[htb]
\centering
\includegraphics[scale=0.7]{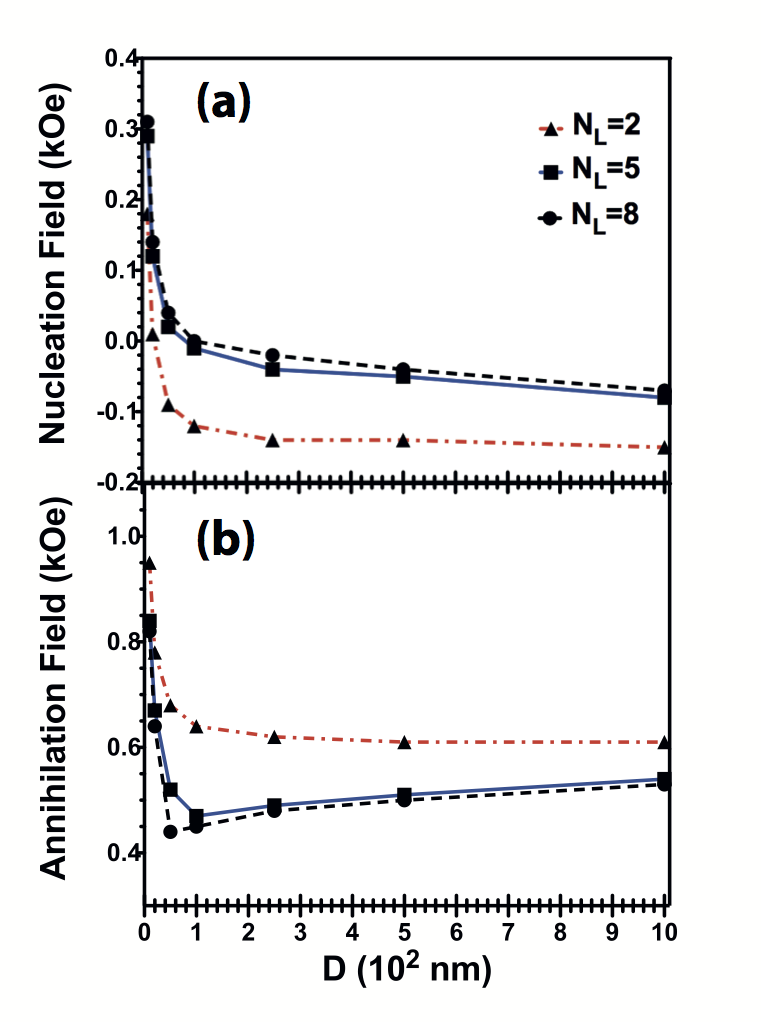}
\caption{(Color online) The figure shows (a) the nucleation and (b) annihilation fields as a function of the distance between the layers of the system at $N_L=2$ (triangle points), $5$ (square points), and $8$ (circle points).}
\label{fig3}
\end{figure}

For these systems, we will focus on the nucleation and annihilation fields of the hysteresis curves. The nucleation (annihilation) field is the magnetic field in which the magnetic reversion of the system starts (finishes). For simplicity we consider the nucleation and the annihilation fields with positive or negative values, depending on the sign of the external applied field for the second or the return branch of the hysteresis loop. The hysteresis curves in figure \ref{fig2} correspond to magnetic multisegmented nanowires with different layers and distances between them. Figure \ref{fig2}(a) shows the hysteresis curves for $N_L=2$. In this figure we observe that the nucleation field decreases when the distance between layers increases. Also we see that the annihilation field decreases when $D$ increases. In both cases we have a monotonic behavior between the nucleation (annihilation) fields with $D$. In figure \ref{fig2}(b), hysteresis curves for $N_L=5$, we observe that the nucleation field has a monotonic behavior with the distance between layers, but the annihilation field does not have a monotonic behavior with $D$ when we observe $D=50$ nm and specially for $D=100$ nm. Also in $D=50$ nm and $D=100$ nm, we observe a second nucleation field at $H=0.18$ kOe ($D=50$ nm) and  $H=0.24$ kOe ($D=100$ nm). Figure \ref{fig2}(c) shows the hysteresis loops for $N_L=8$ layers at different distances. We observe a similar behavior with the figure \ref{fig2}(b). There is a non-monotonic behavior of the annihilation field with the distance between layers, and the hysteresis loops also present a second nucleation field. A summary of the nucleation and annihilation fields is illustrated in figure \ref{fig3}. In figure \ref{fig3}(a) we observe that the nucleation field decreases when the number of layer decreases or the distance between them increases. On the annihilation field, figure \ref{fig3}(b) shows a minimum at $N_L=5$ with $D=100$ nm and $N_L=8$ with $D=50$ nm, but for $N_L=2$ there is not a minimum. We also see that the annihilation field decreases when we increase the number of layers for the system.   

\begin{figure}[htb]
\centering
\includegraphics[scale=0.4]{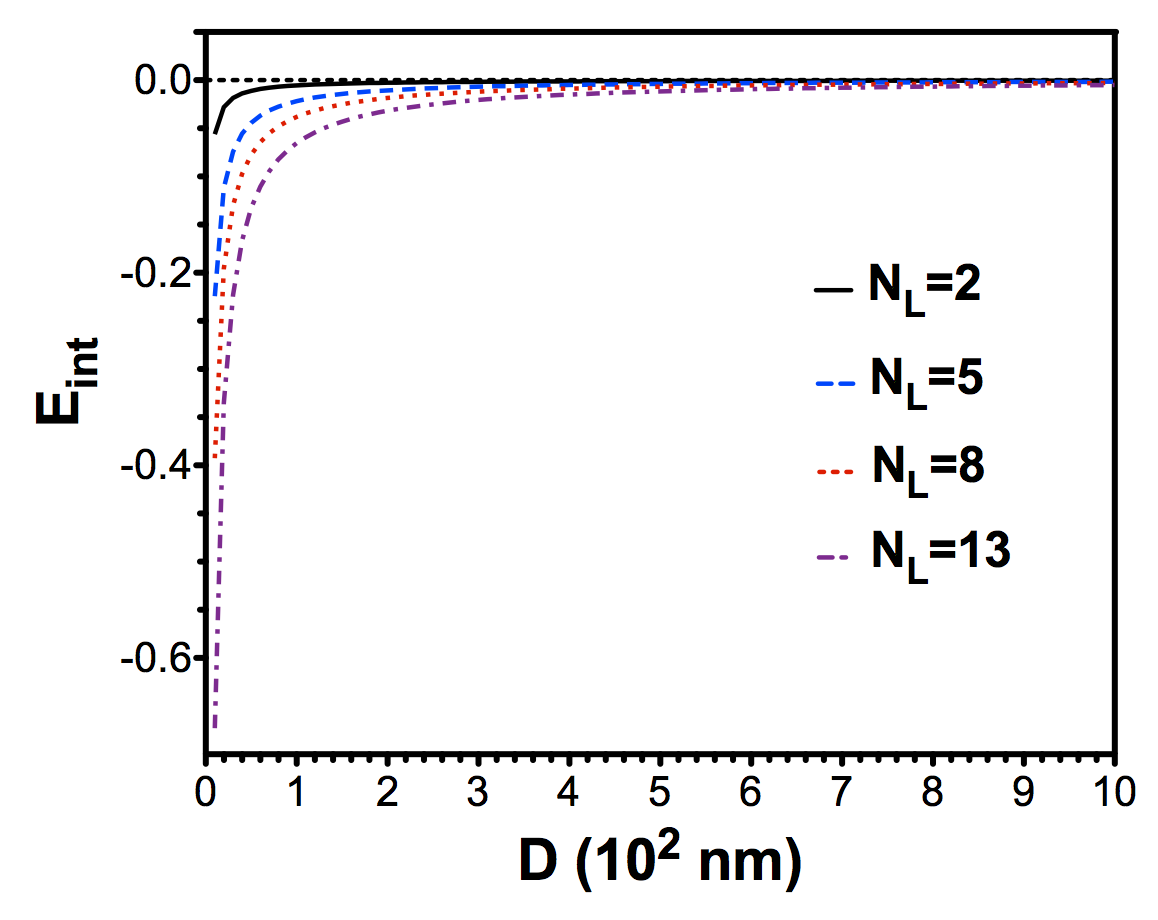}
\caption{(Color online) The normalized magnetostatic interaction among 2, 5, 8 and 13 segments that is in one alumina pore when they have the same magnetic configuration $\vec{\sigma}_i=\hat{z}$.}
\label{fig4}
\end{figure}

\begin{figure}[htb]
\centering
\includegraphics[scale=0.3]{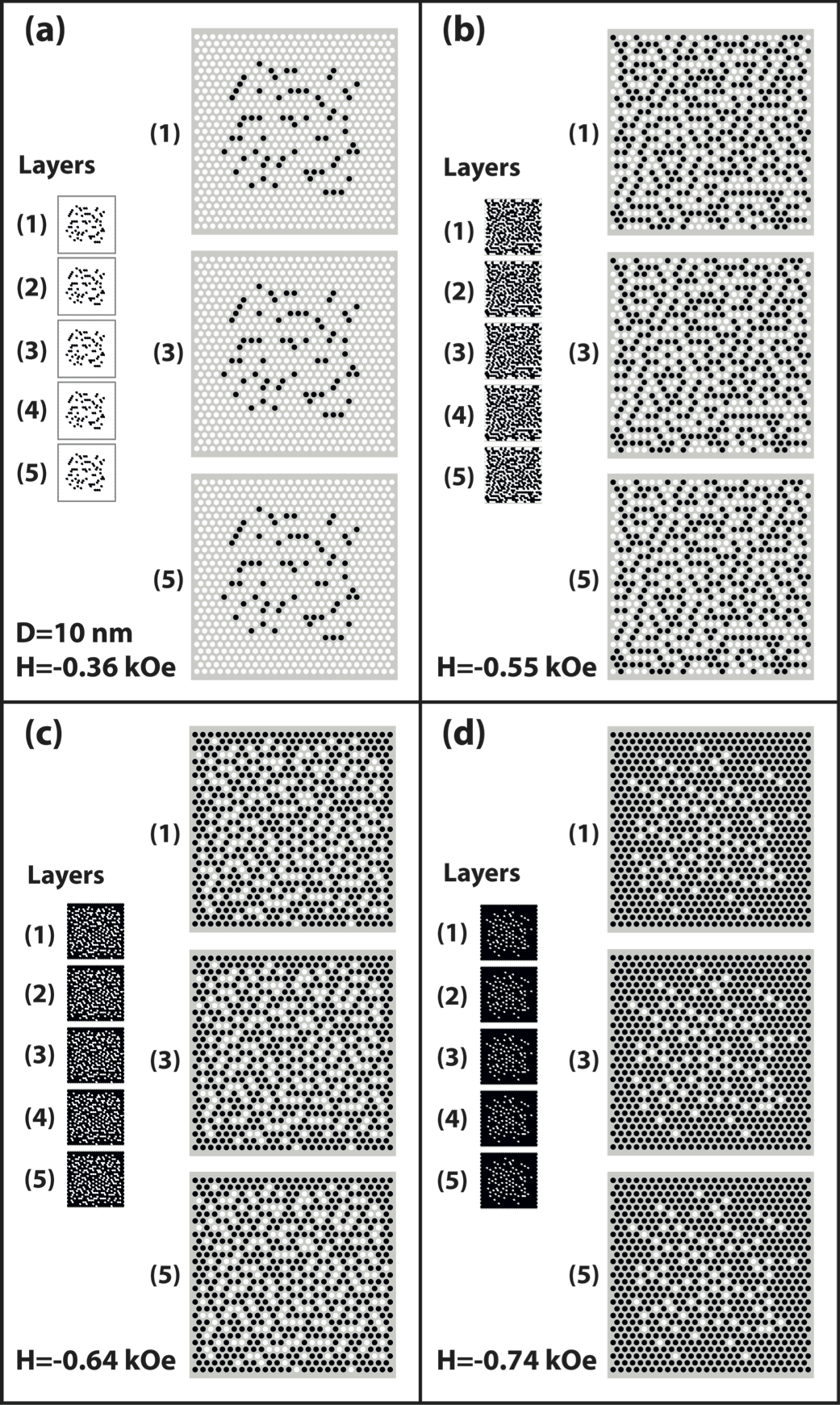}
\caption{Magnetic pattern configuration for $D=10$ nm and $N_L=5$ at different external applied fields from the first branch of the hysteresis curve. In every field,  the figures depict three layers out of five, which correspond to layers one, three, and five. Also there is a small inset of the five layers together. The white circles are magnetic segments that are not reverting their magnetization yet. The black circles means segments that reverted its magnetization.}
\label{fig5}
\end{figure}

To understand the magnetic behavior for the nucleation field, we need to see the magnetostatic interaction among the segments that are present in one alumina pore. Figure \ref{fig4} shows the normalized magnetostatic interaction, $E_{\text{int}}$, among the segments inside in one pore when they have the same magnetic configuration $\vec{\sigma}_i=\hat{z}$. We observe that $E_{\text{int}}$ is negative and decreases, or the system is more stable, when the layer number increases or $D$ decreases. The reduced remanence, or the remanence divided by the saturation magnetization, has the same behavior that the nucleation field, this means it increases its value until one when decreasing $D$ or increasing $N_L$. When we have one single nanowire, the reduced remanence of its hysteresis curve is one. When we have several nanowires in one array, the magnetostatic interaction decreases the reduced remanence to a value less than one due to the spatial distribution on a plane (XY) and the magnetic configuration of the nanowires. This is because the magnetic nanowires are energetically unstable when they are saturated. But if we add more nanowires distributed along $\hat{z}$ (not in the plane (XY)), for instance, other layers forming a multisegmented system, then the magnetostatic interaction decreases until negative values helping to the stability of the nanowires, which means the reduced remanence must increase. In addition, if we have two o more layers, and then we drecrease $D$, the magnetostatic interaction decreases until negative values or the reduced remanence increases until one. The last observation explains the nucleation field and the reduced remanence behavior when D decreases or $N_L$ increases, but this argument does not explain why we observe a non-monotonic behavior in the annihilation fields.

\begin{figure}[htb]
\centering
\includegraphics[scale=0.3]{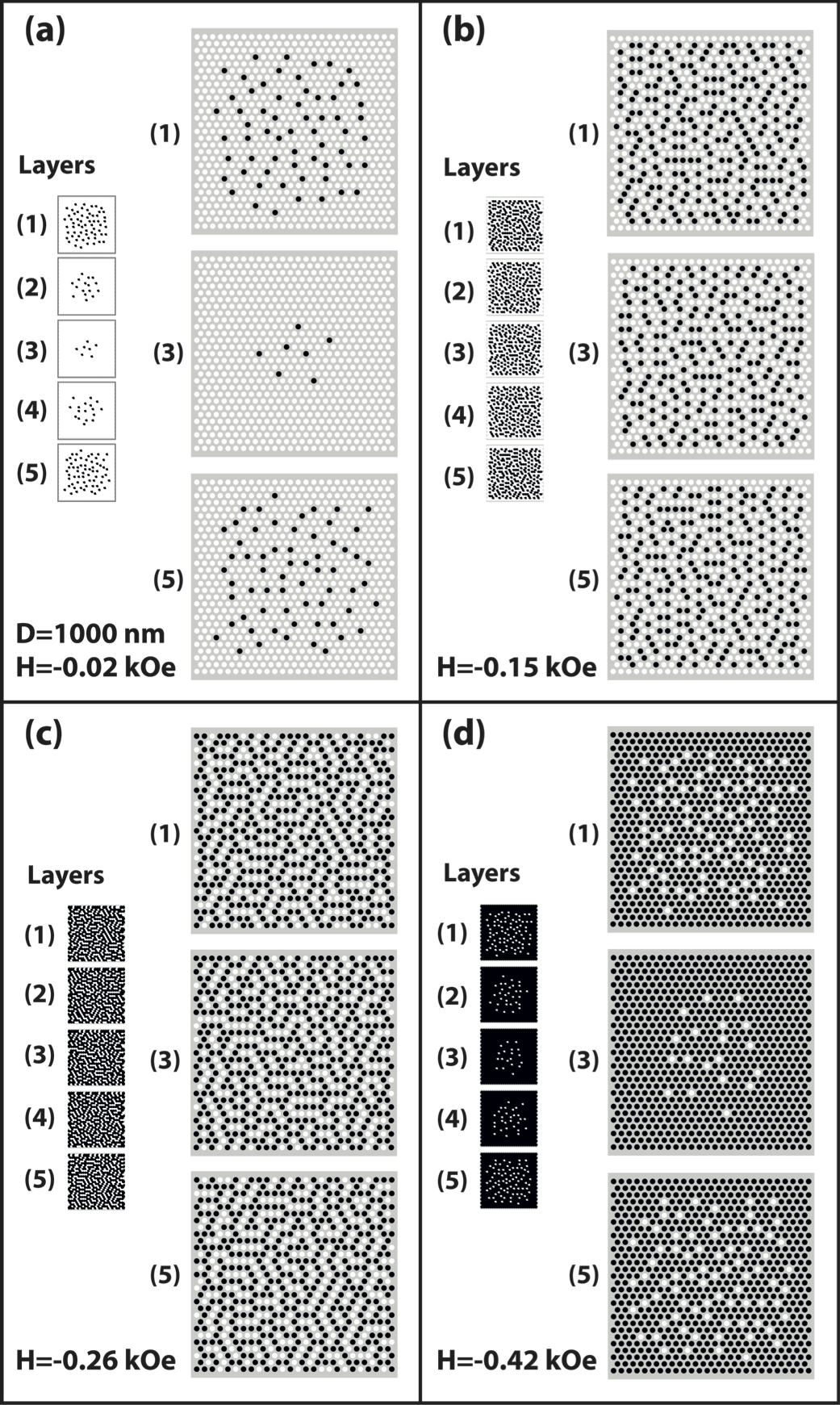}
\caption{Magnetic pattern configuration for $D=1000$ nm and $N_L=5$ at different external applied fields from the first branch of the hysteresis curve. In every field, the figures depict three layers out of five, which correspond to layers one, three, and five. Also there is a small inset of the five layers together. The white circles are magnetic segments that are not reverting their magnetization yet. The black circles means segments that reverted its magnetization.}
\label{fig6}
\end{figure}

%\noindent 
\begin{figure}[htb]
\centering
\includegraphics[scale=0.3]{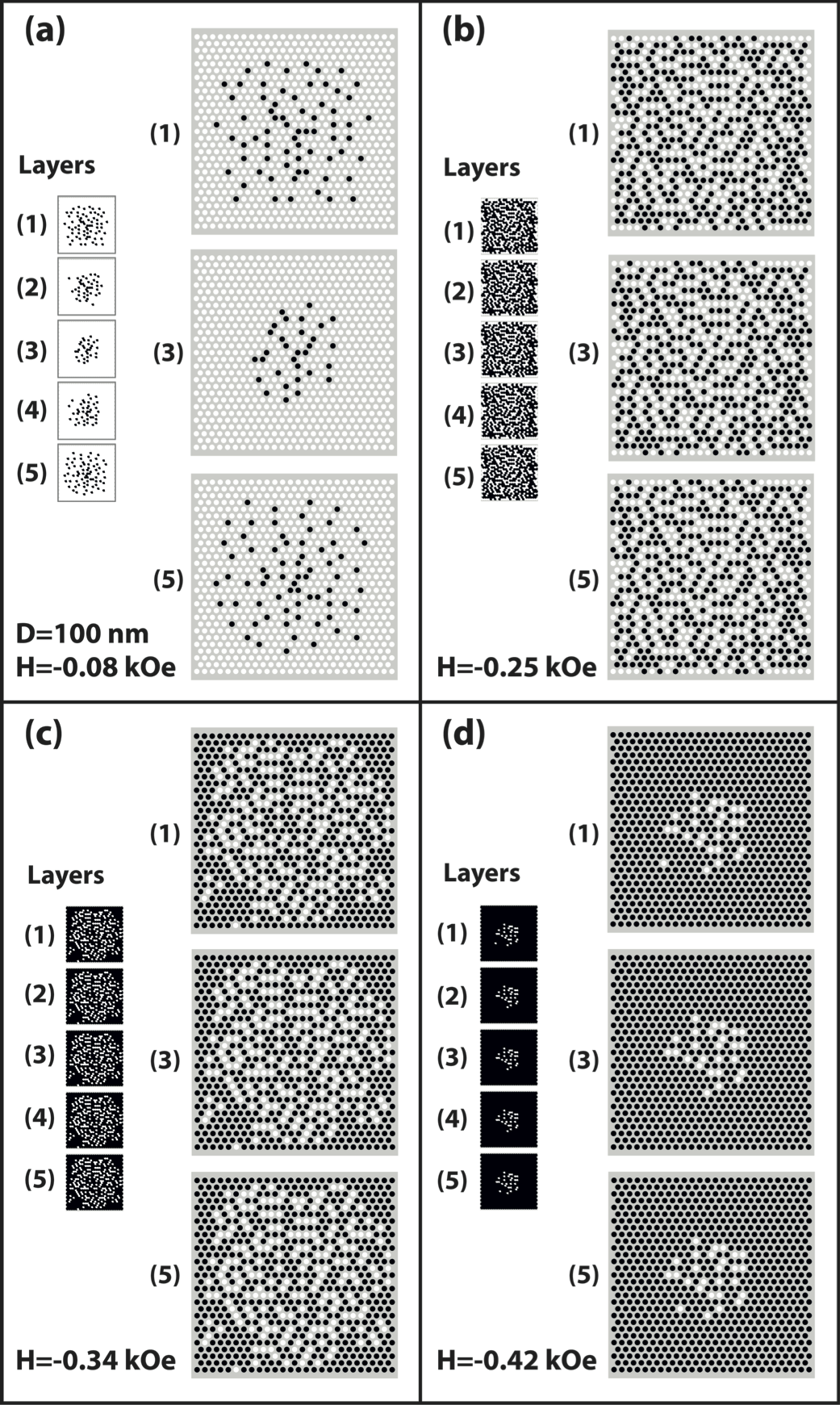}
\caption{Magnetic pattern configuration for $D=100$ nm and $N_L=5$ at different external applied fields from the first branch of the hysteresis curve. In every field,  the figures depict three layers out of five, which correspond to layers one, three, and five. Also there is a small inset of the five layers together. The white circles are magnetic segments that are not reverting their magnetization yet. The black circles means segments that reverted its magnetization.}
\label{fig7}
\end{figure}

With the purpose to explain the non-monotonic behavior of the annihilation fields, we need to understand why the second nucleation fields for the hysteresis curves occur. This is because these second nucleation fields show a change of the magnetic reversions that affect the annihilation fields. This explanation can be obtained by looking how the magnetic multisegmented nanowires revert in these systems. For illustrative purposes, we will show the magnetic reversion for $N_L=5$ at different distances (figures \ref{fig5}, \ref{fig6}, and \ref{fig7}). For each distance we will show four specific applied fields of the magnetic reversion (first branch of the hysteresis loop), where the magnetic reversion of the layer one, three, and five can be observed in these four fields. Also we include an inset of the five layers together with the purpose to have a view of where the magnetic system starts and finishes its magnetic reversion. For each layer we have 740 circles that are mixing in white or black circles. The white circles are magnetic segments with a magnetic state $\vec{\sigma}_i=\hat{z}$, in other words, they are not reverting their magnetization yet. The black circles are magnetic segments with state  $\vec{\sigma}_i=-\hat{z}$, it means that they reverted its magnetization. We will start when the layers have strong magnetostatic interaction ($D=10$ nm), see figure \ref{fig5}. The magnetic system reversion starts in the center of every layer, figure  \ref{fig5}(a). The segments that are in the center of every layer are more energetically unstable than the segments of the border of every layer\cite{escrig1,Raposo}, then part of the segments from the center of every layer revert its magnetization forming antiparallel states with the neighbor segments that are not reverting its magnetization yet. These antiparallel states are energetically stable configurations (negative magnetostatic interaction)\cite{escrig1,Raposo}. After that, when we increase the magnitude of the applied field, the magnetic reversion of the system propagates to the edge of every layer forming antiparallel states, figure \ref{fig5}(b) and figure \ref{fig5}(c). Due to the strong negative magnetostatic interaction among the segments in one pore (see figure \ref{fig4}), the magnetization of the segments in one pore always want to have the same magnetic state \cite{escrig1,morales}, then the system always reverts with the same patterns in all five layers. The five segments in one pore are acting like a single segment, then if we choose a pore $i$, all segments from this pore have a strong magnetic correlation among them. It means all the segments in one pore have the magnetic state $+\hat{z}$ or $-\hat{z}$. Finally the system finishes its magnetic reversion when the segments from the center of every layer revert its magnetization, figure \ref{fig5}(d). This is because these segments have more antiparallel states that the segments from the edge of every layer\cite{escrig1,Raposo}. On the other hand, figure \ref{fig6} shows the case where the layers are approximately non interacting by magnetostatic interaction, $D=1000$ nm. In figure \ref{fig6}(a), the magnetic reversion starts in the center of the layers because the segments are more energetically unstable than the segments from the border of the layers (positive magnetostatic interaction), then part of the segments from the center of the layers revert its magnetization forming antiparallel states with the neighbor segments that are not reverting its magnetization yet\cite{escrig1,Raposo}. These antiparallel states are energetically stable configurations, see figure \ref{fig6}(a). The reason that magnetic segments from the extremity layers (layers 1 and 5) start to revert in contrast with the middle layers (layer 3) is due to the magnetostatic interaction of the segments in one pore. The more stable segments in one pore, when the magnetic reversion of the system starts, are from the middle layers because they have more first neighbors than the segments from the extremities\cite{escrig1,morales}. This means that the segments from the middle layers have stronger negative magnetic interaction than the extremities segments due to the spatial distribution and magnetic configuration of them\cite{escrig1,morales}. So the segments from the extremity layers start to revert first.  When we increase the magnitude of the applied field, the magnetic reversion of the system propagates to the edge of every layer forming antiparallel states, figure \ref{fig6}(b) and figure \ref{fig6}(c). We observe from figure \ref{fig6}(c) that there are not equal magnetic layer patterns when the system reverts. In this system, all segments inside of a pore $i$ has different magnetic states (up or down). This magnetic configuration of the segments in one pore is energetically unstable due to the spatial distribution of the segment in one pore and the different magnetic states of the segments from this pore\cite{escrig1,morales}. The segments want to have all the same magnetic state with this spatial distribution. In addition, at $H=-0.26$ kOe, figure \ref{fig6}(c), every layer has several antiparallel states, where segments that are not reverting its magnetization are forming more antiparallel states from the center than the edge of the layers. Then if we increase the magnitude of the applied field from $H=-0.26$ kOe, the segments from the middle of the layers (layer 3) and the edge of every layers start to revert until that the magnetic reversion of the system finishes in the center of the extremity layers (layer 1 and 5), see figure \ref{fig6}(d). The last part of the magnetic reversion destroys the labyrinth magnetic patterns of figure \ref{fig6}(c). In an intermediate distance, $D=100$ nm, the negative magnetic interaction among the segments from one pore is stronger than $D=1000$ nm and less than $D=10$ nm (figure \ref{fig4}). From figure \ref{fig7}(a), the magnetic reversion starts similarly to D=1000 nm, in other words, the layers do not have the same magnetic patterns and the magnetic reversion starts from the extremity layers (one and five). This is because segments from the extremities, in the same pore, have low negative magnetic interaction between them, so they can have different magnetic state, that is to say, they do not have magnetic correlation between them. It is due to the segment that is in the layer one has low interaction with the segment in the layer five, then they do not have magnetic correlation between them. But for some field, $H=-0.25$ kOe, all the layers have the same magnetic patterns. To understand this situation, we need to consider the five segments that are in the same pore. Although the segments of layer one and layer five do not have a strong magnetostatic interaction, the segments of layer one and layer two have a strong magnetostatic interaction (first neighbors), so that the magnetic reversion of the segments, inside of one pore, occurs in a sequential form from one extremely layer to the other extremely layer, where this sequence happens at different nearby fields, that is to say, after reverting the magnetization of the segment in the layer one, the segment of layer two reverts at some nearby applied field, and so on with the segments in the layer three, four, and five. The same situation happens when the reversion starts with the segment of layer 5. When the segment in the layer five reverts its magnetization, the segment in the layer four is the following to revert its magnetization, and so on with the segments in the layer three, two, and one. Thus, for a given applied field, the system will have all their five layers with the same magnetic patterns. From the external field $H=-0.25$ kOe to $H=-0.34$ kOe, the system reverts like the system with $D=10$ nm (all layers having the same magnetic pattern). Then the non-monotonic behavior of the annihilation field and the second nucleation field is due to the change of the magnetic reversion from a system that has non magnetostatic interaction among the layers to a system that has a very strong magnetostatic interaction among the layers. Finally the magnetic system loses this behavior of the strong magnetostatic interaction after the external magnetic field $H=-0.34$ kOe, like figure \ref{fig7}(d).

\begin{figure}[htb]
\centering
\includegraphics[scale=0.25]{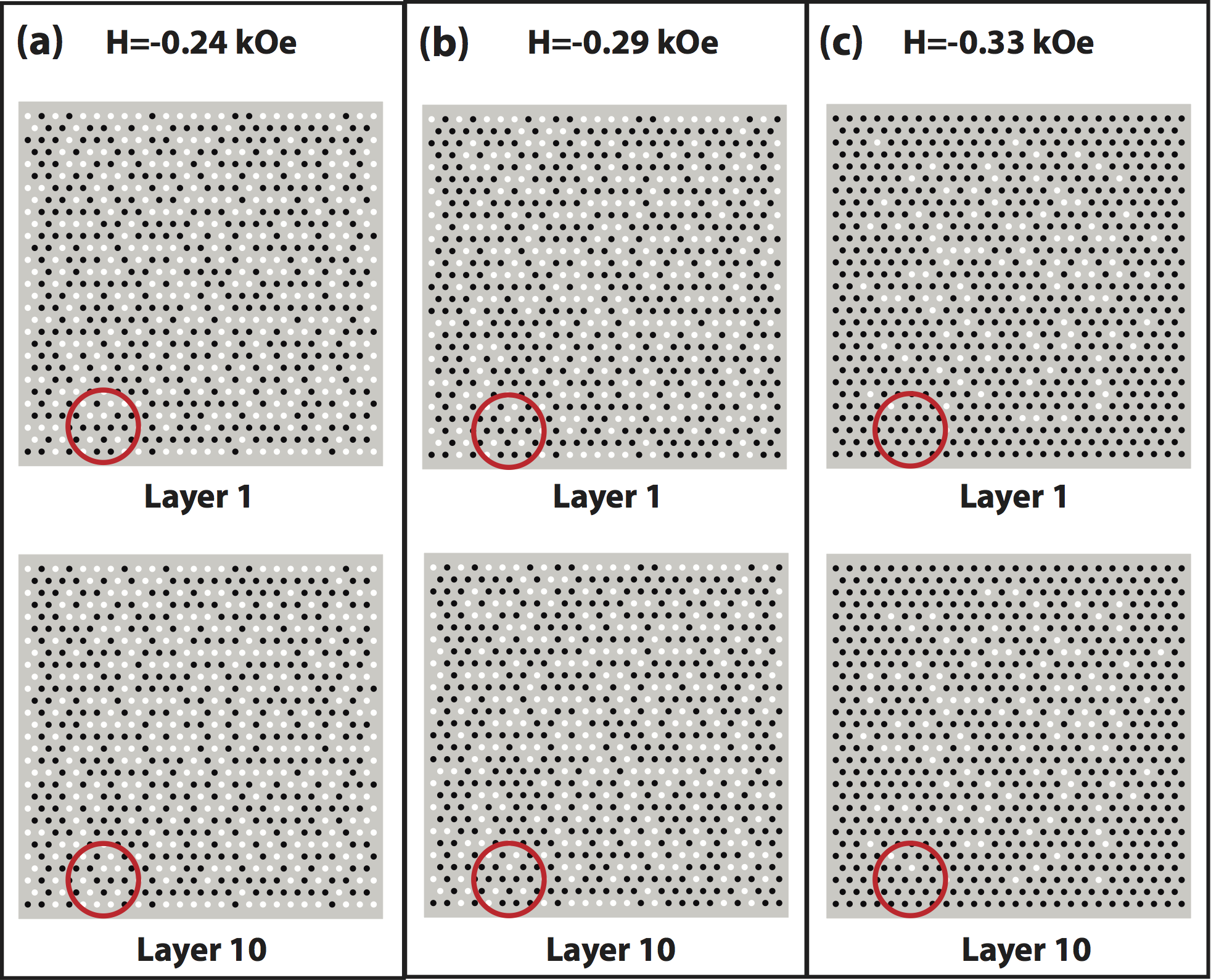}
\caption{(Color online) Magnetic pattern configuration of the extremity layers (layers one and ten) for $D=100$ nm and $N_L=10$ at different external applied fields from the first branch of the hysteresis curve. The magnetic correlations are $\mathcal{C}=0.90$ (a), $\mathcal{C}=1.00$ (b), and $\mathcal{C}=0.97$ (c). The red big circle help to compare the magnetic pattern of the layers one and ten. }
\label{fig8}
\end{figure}

\begin{figure}[htb]
\centering
\includegraphics[scale=0.4]{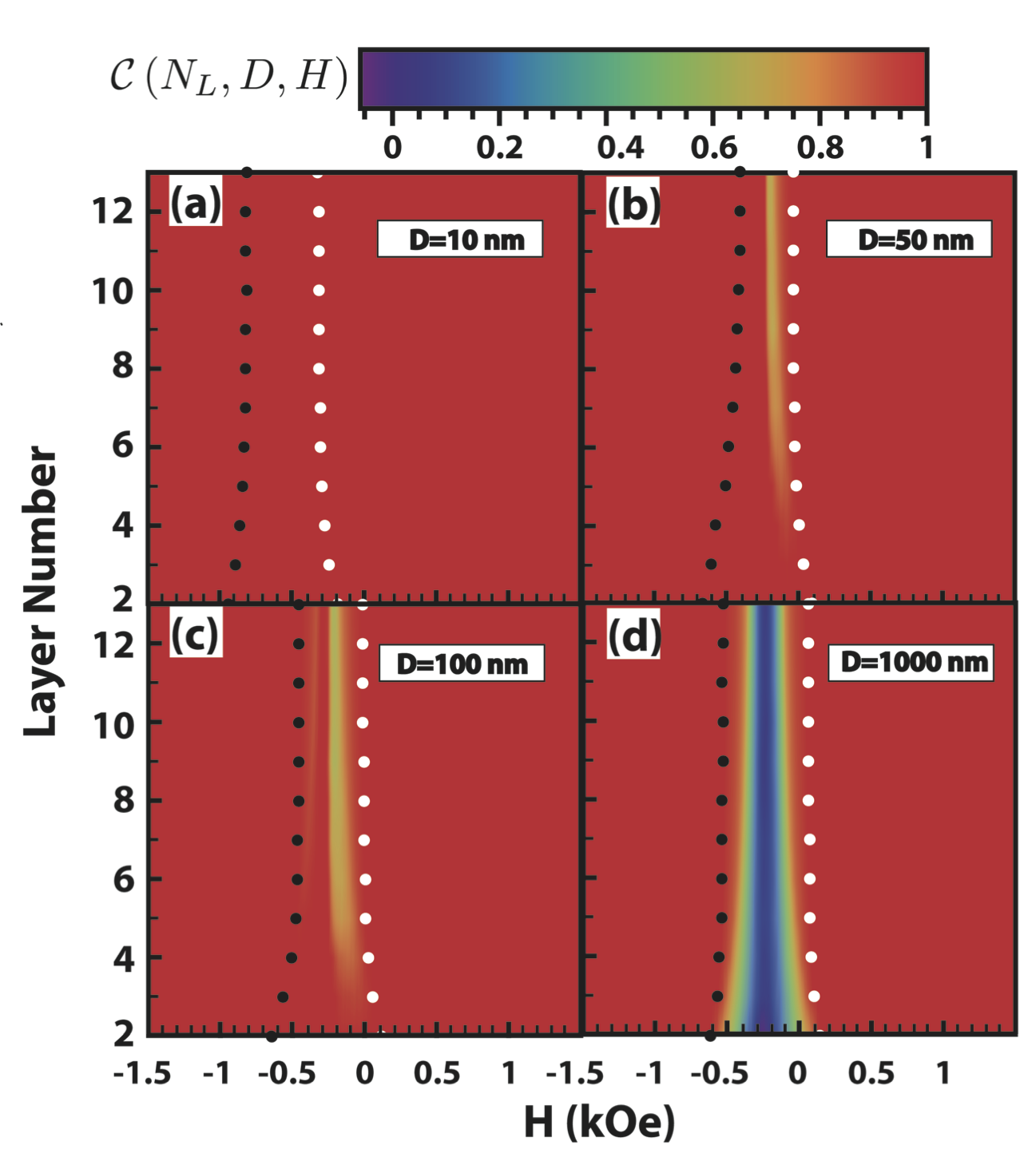}
\caption{(Color online) Magnetic correlation among the layers of the system at  $D=10$ nm, $D=50$ nm, $D=100$ nm, and $D=1000$ nm. The red color means strong magnetic correlation, blue color means non magnetic correlations (for more detail see the color bar). The correlation values are from the first branch of the hysteresis curve. The white (black) dots correspond to where the system starts (finishes) its magnetic reversion. The correlation function from the right (left) side of the white (black) dots are always $\mathcal{C}=1$ due to that the systems are saturated in $+\hat{z}$ ($-\hat{z}$).}
\label{fig9}
\end{figure}

With the purpose to resume and clarify the reversal modes shown in figures \ref{fig5}, \ref{fig6}, and \ref{fig7}, we will study the magnetic correlation among the layers. The magnetic correlation of the magnetic patterns among the layers, when the magnetic system is reverting, can be quantified in terms of the layer number, distance between the layers, and the external field. We define the following magnetic correlation function, $\mathcal{C}$, as

 \begin{equation}
\mathcal{C}\left(N_L,D,H\right)=\sum_{\substack{
  i=1\\
  j=1\\
i \neq j}}^{N_L} \sum_{k=1}^{N_P}  \frac{\vec{\sigma}\left(x_k,y_k,z_i,H\right) \cdot \vec{\sigma}\left(x_k,y_k,z_j,H\right)}{N_P N_L \left(N_L-1\right) }
\label{Heff}
\end{equation}

where $x_k$ and $y_k$ represent the position of the pore $k$ in one layer, and $z_i$ represents the position of the layer $i$. This magnetic correlation function can take values from $-1$ to $1$. The value $\mathcal{C}=1$ represents all the layers which have the same magnetic patterns, for instance, the saturation state has $\mathcal{C}=1$. The value $-1$ corresponds to the layers which have opposite magnetization, for instance, a magnetic system with $N_L=2$ where the first layer has all the segments with magnetic states up and the second layer all the magnetic segments have the magnetic state down. The value $0$ means that the layers do not have magnetic correlation among them. Figure \ref{fig8} helps to visualize equation \ref{Heff}. This figure shows the magnetic pattern configuration of the extremity layers (layers one and ten) for $D = 100$ nm and $N_L = 10$ at different external applied fields from the first branch of the hysteresis curve. In this figure, the red big circle helps to compare the magnetic pattern configuration of the layers. For this system the magnetic correlation is equal to one at $H=-0.29$ kOe (figure \ref{fig8}(b)), but less than one at $H=-0.24$ kOe (figure \ref{fig8}(a)) and $H=-0.33$ kOe (figure \ref{fig8}(c)). Then if we compare the magnetic pattern configuration of the layers one and ten from figure \ref{fig8}(b), we can observe that they have the same magnetic configuration, but it is not the case for the other fields (figures  \ref{fig8}(a) and  \ref{fig8}(b)). Figure \ref{fig9} illustrates the magnetic correlation as a function of the applied field and the layer number of the systems at different $D$. The red color means strong magnetic correlation, blue color means non magnetic correlations, see color bar in figure \ref{fig9}. The correlation values are from the first branch of the hysteresis curve. The white (black) dots correspond to where the system starts (finishes) its magnetic reversion. The correlation function from the right (left) side of the white (black) dots are always $\mathcal{C}=1$ due to the fact that the systems are saturated in $+\hat{z}$ ($-\hat{z}$). From figure \ref{fig9}, $D= 10$ nm, we observe that the magnetic reversion of these systems always has the same magnetic patterns in all the layers. It is due to the strong magnetostatic interaction among the segments in one pore. This reversal mode has a positive strong magnetic correlation among the layers with $\mathcal{C}=1$. When we increase the distance between the layers, $D=50$ nm, the magnetostatic interaction among the first neighbor segments are strong, but it is weak for the second or high order neighbors. Then we observe that approximately in the middle of the reversion, the magnetic correlation increases its value until one. A similar situation happened for $D=100$ nm. In both cases we see a reversal mode with positive weak magnetic correlation among the layers. For distance $D=1000$ nm, the segments from different layers have approximately non interaction among them. Then we observe how the magnetic correlation decreases from one to zero and after that increases its value to one. Then we observe that this reversal mode does not have magnetic correlation among the layers. To finalize, in a real system there is a small deviation in lengths and diameters of the magnetic segments, producing a distribution of the anisotropy field and switching field values for each segment. These distributions make small changes of the values of the nucleation and annihilation fields, but still can be observed the additional nucleation field and the non-monotonic behaviors of the annihilation field because it depends on how strong the magnetostatic interaction among the layers are.

\section{Conclusion}

In conclusion, in the present work a detailed investigation of the magnetic reversal process in multisegmented nanowires with non-magnetic spacer and long aspect ratio has been carried out. By means of Monte Carlo simulation we studied the nucleation and annihilation fields in different hysteresis curves. The nucleation fields increase when the number of layer increases or the length of the non-magnetic spacer decreases. The annihilation fields have non-monotonic behavior when varying $D$ for certain $N_L$. This non-monotonic behavior is due to the different reversal modes that depend on the number of magnetic segments by pores, as well as the spacing between the segments of the same pore. For these systems we observed three magnetic reversal modes depending on the behavior of the magnetic correlation function with the applied field. One magnetic reversal mode has the magnetic correlation function equal to one for every applied field. This mode has a positive strong magnetic correlation among the layers, and this means that every layer has the same magnetic patterns when the system reverts its magnetization. The second reversal mode is observed when the magnetic correlation function only takes the value one at some range of the applied field. This means that the magnetic patterns of every layer are equal at some specific range of the applied field. In this mode the correlation function is always greater than zero. Then, this mode has a positive weak magnetic correlation among the layers. The last reversal mode occurs when the magnetic correlation function has value zero at some field and value one only in the saturated cases. This mode does not have magnetic correlation among the layers. This means that the magnetic patterns of every layer are always different. Thus, these results might allow the production of magnetic devices with multisegmented nanowire arrays with non-magnetic spacer.  For instance, the knowledge of the different reversion of the magnetization for multisegmented nanowires gives a good guideline for experimentalists who want to use multisegmented nanowires for storage devices. In addition, the reversal mode with positive weak magnetic correlation among the layers presents the characteristic that the system has the same magnetic pattern in all the layers only at some specific fields, and then if we measure the spatial distribution of the magnetization in both extremity layers and compare these measurements (e.g. see figure \ref{fig8}), we can build a sensor field. Also this result can be used in cryptography, for instance, it opens an access only when we applied a specific magnetic field (trigger field devices).

\begin{acknowledgements}
 We acknowledge financial support from FONDECYT 11121214 and 1120356, Financiamiento Basal para Centros Cient\'ificos y Tecnol\'{o}gicos de Excelencia FB 0807,  Concurso Inserci\'{o}n en la Academia-Folio  791220017.
\end{acknowledgements}

\end{document}